\documentstyle[twocolumn,aps,prl,epsf]{revtex}
\begin{document}
\newcommand{\beq}{\begin{equation}}
\newcommand{\eeq}{\end{equation}}

\title{Nonlinear AC resistivity in $s$-wave and $d$-wave disordered granular
 superconductors}

\author{Mai Suan Li$^1$, Hoang Zung$^2$ and D. Dom\'{\i}nguez$^3$}

\address{
$^1$Institute of Physics, Polish Academy of Sciences,
Al. Lotnikow 32/46, 02-668 Warsaw, Poland\\
$^2$Vietnam National University, Ho Chi Minh City\\
$^3$ Centro At\'{o}mico Bariloche, 8400 S. C. de Bariloche, Rio Negro,
Argentina}

\address{
\centering{
\medskip\em
{}~\\
\begin{minipage}{14cm}
We model  $s$-wave and $d$-wave disordered granular superconductors 
with a three-dimensional lattice of randomly distributed
Josephson junctions with finite self-inductance. 
The nonlinear ac resistivity $\rho_2$
of these systems was calculated 
using Langevin dynamical equations. 
The current amplitude dependence of $\rho_2$
at the peak position is found to be a power law characterized
by exponent $\alpha$.  The later is not universal but depends on the 
self-inductance and current regimes. 
In the weak current regime $\alpha$ is independent of the self-inductance
and $\alpha = 0.5 \pm 0.1$
for both of $s$- and $d$-wave materials. In the strong current regime
the values of $\alpha$  depend on the screening.
We find $\alpha \approx 1$ for some interval of inductance  which agrees
with the experimental finding for $d$-wave ceramic superconductors.
{}~\\
{}~\\
{\noindent PACS numbers: 75.40.Gb, 74.72.-h}
\end{minipage}
}}

\maketitle




The symmetry of the superconducting pairing function has been of 
great interest lately. The gap of conventional superconductors 
has $s$-wave symmetry whereas there is now good evidence that
the superconducting gap of the high-$T_c$ cuprates has
$d$-wave symmetry \cite{HarlKirt}.
Granular superconductors are usually described as 
a random network of superconducting grains  coupled by Josephson
links \cite{ebner,Joseph2}. In high-$T_c$ ceramics,
depending on the relative orientation of the $d$-wave 
superconducting grains, it is possible
to have  weak links with negative Josephson coupling \cite{sigrist}, 
which are called $\pi$-junctions. 
The existence of these $\pi$-junctions may cause, e.g.,
the paramagnetic Meissner effect \cite{sigrist} observed at low
magnetic fields \cite{pme_exp}.

Recently, Kawamura \cite{Kawamura95} proposed that a novel 
thermodynamic phase may occur 
in zero  external magnetic field in unconventional superconductors.  
This phase is characterized by a broken time-reversal symmetry and is called
chiral glass phase.
The frustration effect due to the random distribution of $\pi$ junctions
leads to a glass state of quenched-in ``chiralities'', which are
local loop supercurrents circulating over grains and
carrying a half-quantum of flux.\cite{KawLi97}
Evidence for the transition to chiral glass has been seen from experimental
studies of the nonlinear ac magnetic susceptibility \cite{Matsuura1},
the dynamic scaling \cite{Papadopoulou1} and the aging phenomenon
\cite{Papadopoulou2}. The susceptibility measurements of Ishida {\em et al.}
\cite{Ishida} do not, however, support the existence of the chiral glass. 

In order to further probe existence of the chiral glass phase
Yamao {\em et al.}\cite{Matsuura} have measured
the ac linear resistivity $\rho _0$ and the nonlinear resistivity $\rho_2$
of ceramic superconductor YBa$_2$Cu$_4$O$_8$. 
$\rho_2$ is
defined  as the third coefficient of the expansion
of the voltage $V(t)$ in terms of the external current $I_{ext}(t)$:
\begin{equation}
V \; \; = \; \rho_0 I_{ext} + \rho_2 I_{ext}^3 + ... \; \; .
\end{equation}
When the sample is driven by an ac current
 $I_{ext}(t) = I_0 \sin(\omega t)$, one can relate
$\rho _2$ to  third harmonics
$V'_{3\omega}$ in the following way
\begin{eqnarray}
\rho_2 \; \; = \; \; -4 V'_{3\omega}/I_0^3 \; , \nonumber\\
V'_{3\omega} \; \; = \; \; \frac{1}{2\pi}
\int_{-\pi}^{\pi} \; V(t) \sin (3\omega t) 
d(\omega t) \; \; .
\end{eqnarray}

Yamao {\em et al.} have made two key observations. First, since the
linear resistivity does not vanish at the peak position of $\rho_2$ they 
identify
the transition as a transition to the chiral glass phase. 
Second interesting observation is
the power law dependence
of $|V'_{3\omega}(T_p)/I_0)^3|$ (or of  $\rho_2$)
at its maximum position $T_p$
on $I_0$: $|V'_{3\omega}(T_p)/I_0^3| \sim I_0^{-\alpha}$. 
The experimental value of the power law exponent was
$\alpha \approx 1.1$.
Using the XY-like model for $d$-wave superconductors Li and Dominguez 
\cite{LiDomin} were able to
reproduce the experimental results of Yamao {\em et al.}\cite{Matsuura}
qualitatively. The quantitative agreement was, however, poor and the role of
inductance was not explored. Namely, 
$\alpha$ was computed only for one value of dimensionless inductance
$\tilde{L}$=1 and with large error bars \cite{LiDomin}:
$\alpha = 1.1 \pm 0.6$.

The goal of this paper is twofold.
First, we calculate $\alpha$ with high accuracy for both of $s$- and $d$-wave
systems using  the Langevin equations for the XY-like model with screening.
Second, we try to answer the question if it is possible to discriminate between
$s$- and $d$- pairing symmetry by measurements of $\alpha$.
We show that there are two distinct current regimes for $\alpha$.
In the weak current regime (WCR) (small $I_0$) this exponent does not depend 
on the inductance 
and $\alpha = 0.50 \pm 0.1$ for $s$- and $d$-wave ceramics.
In the strong current regime (SCR) $\alpha$ depends on the
screening.
For small $\tilde{L}$ we obtain $\alpha_{d{-wave}}
>\alpha_{s{-wave}}$, possibly because
in the weak screening limit the energy landscape of the $d$-wave
case is more rugged than the $s$-wave case.  
As the self-inductance increases the number of energy local minima gets
smaller \cite{Li3} and the behavior of the two systems becomes  more
similar, with the values of $\alpha$ being almost the same. 
For the $d$-wave system in the SCR and with $1 < \tilde{L} \leq 5$ 
we find $\alpha \approx 1.0 $ which
agrees with the experimental value \cite{Matsuura}.

We consider the following ``coarse grained" 
Hamiltonian\cite{Dominguez,KawLi,Sasik}
\begin{equation}
{\cal H} = - \sum _{<ij>} J_{ij}\cos (\theta _i-\theta _j-A_{ij})+
\frac {1}{2 L} \sum _p \Phi_p ^2, 
\end{equation}
where $\theta _i$ is the phase of the condensate of the grain
at the $i$-th site of a simple cubic lattice,
$J_{ij}$ denotes the Josephson coupling
between the $i$-th and $j$-th grains,
$L$ is the self-inductance of a loop (an elementary plaquette),
while the mutual inductance between different loops
is neglected.
The first sum is taken over all nearest-neighbor pairs and the
second sum is taken over all elementary plaquettes on the lattice.
Fluctuating  variables to be summed over are the phase variables,
$\theta _i$, at each site and the gauge variables, 
$A_{ij}=\frac{2\pi}{\phi_0} \int_{i}^{j} \, \vec{A}(\vec{r})
d\vec{r}$, at each
link. $\Phi_p = \frac{\phi_0}{2\pi} \sum_{<ij>}^{p} A_{ij}$ 
is the total magnetic flux threading through the
$p$-th plaquette, and $\phi _0$ denotes the flux quantum.
The effect of screening currents inside grains is not considered explicitly, 
since for large length scales they simply lead to a hamiltonian ${\cal H}$
with an effective self-inductance $L$ \cite{Sasik}.

For the $d$-wave superconductors
we assume $J_{ij}$ to be an independent random variable
taking the values $J$ or $-J$ with equal probability ($\pm J$ or bimodal
distribution), each representing 0 and $\pi$ junctions.
For the  $s$-wave supercondutors  $J_{ij}$ is always positive but distributed
uniformly between 0 and 2$J$.
It should be noted that model (3) with uniform couplings was first studied 
by Dasgupta and Halperin \cite{Halperin}.
Random $\pi$-junction models (in which $J_{ij}$ is allowed to take negative  
values with certain probability) have also been
adequate to explain several phenomena observed in high-$T_c$ 
superconductors
such as the anomalous microwave absorption,\cite{Dominguez1,Li3}
the compensation effect \cite{Li}, the effect
of applied electric fields in the apparent critical current \cite{DWJ}
and the aging effect.\cite{LiNord}

In order to study transport properties, we use
the resistively shunted
junction model.\cite{Joseph2}
Then in  addition to the Josephson current one has
 the contribution of a dissipative
ohmic current due to an intergrain resistance $R$ and the Langevin noise
current.
We have redefined notation:
the site of each grain is at position ${\bf n}=(n_x,n_y,n_z)$
(i.e. $i\equiv{\bf n}$); the lattice directions are
${\bbox \mu}={\hat{\bf x}}, {\hat{\bf y}}, {\hat{\bf z}}$;
the link variables are between sites ${\bf n}$ and ${\bf n}+{\bbox \mu}$
(i.e. link $ij$  $\equiv$ link ${\bf n},{\bbox\mu}$);
and the plaquettes $p$ are defined by the site ${\bf n}$ and
the normal direction ${\bbox\mu}$ (i.e plaquette $p$ $\equiv$ plaquette
${\bf n},{\bbox\mu}$, for example the plaquette ${\bf n}, {\hat{\bf z}}$ is
centered at position ${\bf n}+({\hat{\bf x}}+{\hat{\bf y}})/2$).
Then  the gauge invariant phase differences 
$\theta_\mu({\bf n})=\Delta_\mu^{+}\theta({\bf n})-A_\mu({\bf n})$ obey
the following equations \cite{Joseph2,Dominguez}:
\begin{eqnarray}
\frac{\hbar}{2eR}\frac{d\theta_\mu({\bf n})}{dt} \; = \;
-\frac{2e}{\hbar}J_\mu({\bf n})\sin\theta_\mu({\bf n})
-\delta_{\mu,y}I_{ext}\nonumber\\ 
-\frac{\hbar}{2e{L}}\Delta_\nu^{-}\left[
\Delta_\nu^{+}\theta_{\mu}({\bf n})-\Delta_\mu^{+}\theta_\nu({\bf n})\right]
 -\eta_\mu({\bf n},t) \, \; , \nonumber\\
\langle \eta_\mu({\bf n},t) \eta_\mu'({\bf n'},t') \rangle \; \; = \; \; 
\frac{2kT}{R} \delta_{\mu \mu'} \delta_{{\bf n} {\bf n}'} \delta (t-t')
\; \; ,
\label{Langevin}
\end{eqnarray} 
where  
$\eta_\mu({\bf n},t)$ is the Langevin noise current.
The forward difference operator is $\Delta_{\mu}^{+}\theta_\nu({\bf n})=
\theta_\nu({\bf n}+\mu)-\theta_\nu({\bf n})$
and the backward operator  $\Delta_{\mu}^{-}\theta_\nu({\bf n})=\theta_\nu({\bf n})-
\theta_\nu({\bf n}-\mu)$.
In what follows we will consider currents normalized by $I_J=2eJ/\hbar$,
time by $\tau=\phi_0/2\pi JR$, voltages by $RI_J$, 
temperature by $J/k_B$ and inductance by
$\phi_0/2\pi J$.
Free boundary conditions and numerical integration are implemented
in the same way as in \cite{LiDomin,Dominguez}.
Depending on values of $I_0$ and $\omega$ the number of
samples used for the disorder-averaging ranges between 5 and 800. The 
number of integration steps is chosen to be $10^5 - 5\times 10^5$.

\begin{figure}
\epsfxsize=3.2in
\centerline{\epsffile{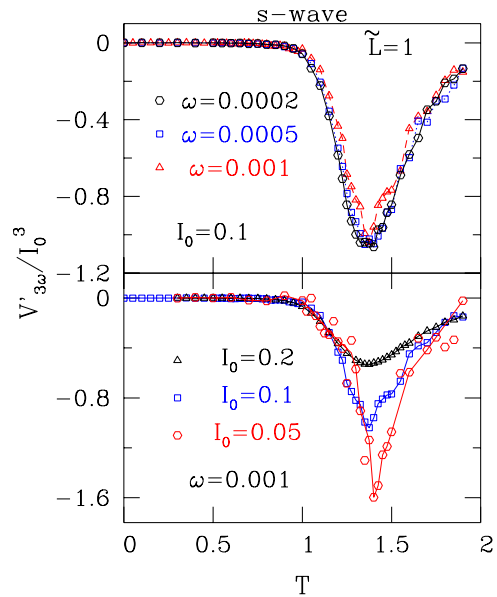}}
\vspace{0.2in}
\caption{ a) Upper panel: the temperature dependence of $V'_{3\omega}/I_0^3$
for the $s$-wave system. System size
$l=8, \tilde{L}=1$ and $I_0 =0.1$.
The open triangles, squares and hexagons correspond to
$\omega = 0.001, 0.0005$ and 0.0002. The peak is located at $T_p=1.4$.
b) lower panel: the same as in upper panel but
$\omega =0.001$.
The open triangles, squares and hexagons correspond to
$I_0 = 0.2, 0.1$ and 0.05.
The results are averaged over
15 - 40  samples.}
\end{figure}

The temperature dependence of the {\em nonlinear} 
resistivity $\rho_2$ of the $s$-wave system
for $I_0=0.1$ and for different values of $\omega$
is shown in upper panel of Fig. 1. Similar to the $d$-wave case \cite{LiDomin}, 
there is no visible dependence on $\omega$. As seen in lower pannel,
as $I_0$ decreases peak values of $\rho_2$ tends to diverge. For
$\tilde{L}=1$ 
the peak is located at $T_p=1.4$ and
it coincides with 
the metal -- superconductor transition 
at which thermodynamic quantities diverge 
and the {\em linear} resistivity $\rho_0$ vanishes. 
It should be noted that our disordered $s$-wave model 
is different from the gauge glass model \cite{Bokil} (in the later case the 
screening spoils the transition to the superconducting state).
Fig. 2 shows the $I_0$ dependence of $max |V'_{3\omega}/I_0^3|$ of
the $s$-wave samples ($\tilde{L}$=1). Clearly, we have two distinct
regimes for small
and large currents.
In the WCR ($I_0 \le 0.1$) $\alpha = 0.50 \pm 0.04$ and
$\alpha=0.51 \pm 0.03$
for $l=8$ and $l=12$, respectively.
In the second regime we obtain $\alpha = 1.0 \pm 0.05$ and
$\alpha=1.07 \pm 0.02$ for $l=8$ and $l=12$, respectively.
Since within the error bars the finite system size effect is negligible,
we will consider only the system size $l=8$.

\begin{figure}
\epsfxsize=3.2in
\centerline{\epsffile{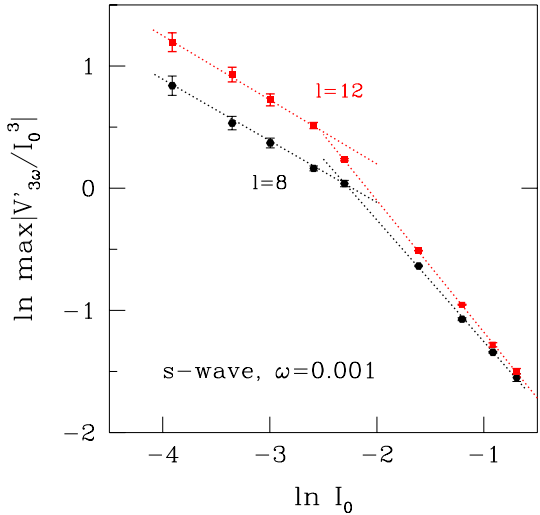}}
\caption{The current dependence of $max |V'_{3\omega}/I_0^3|$ for $s$-wave
superconductors. We choose
$\omega=0.001$ and $\tilde{L}=1$. 
In the WCR $\alpha = 0.50 \pm 0.04$ and $0.51 \pm 0.03$
for the system size  $l=8$ and 12, respectively.
In the SCR $\alpha = 1.0 \pm 0.05$ and
$\alpha=1.07 \pm 0.02$ for $l=8$ and $l=12$, respectively.
The results are averaged over 5 - 800 samples.}
\end{figure}

Fig. 3 shows the dependence of $max |V'_{3\omega}/I_0^3|$ on $I_0$ for
the $d$-wave case ($l=8$ and $\omega=0.001$). In the weak
current part one has
$\alpha=0.51 \pm 0.03$, $0.45 \pm 0.05$, $0.48 \pm 0.05$
and $0.43 \pm 0.06$ for $\tilde{L}$=0.1, 1, 10 and 20, respectively. 
Clearly, within error bars $\alpha$ is not sensitive
to the screening. In the SCR it becomes dependent
on $\tilde{L}$:
$\alpha=1.8 \pm 0.16, 1.56 \pm 0.17, 0.97 \pm 0.02$ and $0.60 \pm 0.02$
for $\tilde{L}$=0.1, 1, 10 and 20, respectively. 
Fig. 4 shows the results obtained in the SCR for
$s$- and $d$-wave systems with different values of $\tilde{L}$.
The power law region of the $d$-wave case is sensitive to
the screening and is narrower than its $s$-wave counterpart.

The dependence of $\alpha$ on $\tilde{L}$ in the SCR is
shown in Fig. 5. Such a dependence may be understood taking into account the
interplay between the thermal fluctuations and the rugged energy landscape.
In the weak screening limit the later plays an important role and $\alpha$
of the $d$-wave system is bigger than that for the $s$-wave one. As $\tilde{L}$
increases the thermal fluctuations take over and the opposite situation
would happen. The pronounced difference between two types of symmetry
is seen only in the weak screening region.

It is tempting to interpret the two regimes for $\alpha$  as
the WCR corresponding to the critical regime for $\rho_2(T_c,I_0)$
(since $I_0\rightarrow0$)
and the SCR corresponding to a mean-field regime
(away from criticality). If there is a continuous phase transition at
a critical temperature $T_c=T_p$, then current-voltage
scaling \cite{ffh} predicts that $V \sim I^{\frac{z+1}{d-1}}$
at $T_c$, with $z$ the dynamical exponent.
Therefore, the non-linear resistivity should be $\rho_2(T_c) \sim
I_0^{\frac{z+1}{d-1}-3}$, and thus the expected WCR value is
$\alpha=(5-z)/2$ in $d=3$. This predicts that
a peak in $\rho_2(T)$ at $T_c$ is possible if $z<5$ (i.e. $\alpha>0$).
In our case, we obtain $\alpha\approx0.5$ and therefore $z\approx4$
for the disordered $s$-wave transition.

\begin{figure}
\epsfxsize=3.2in
\centerline{\epsffile{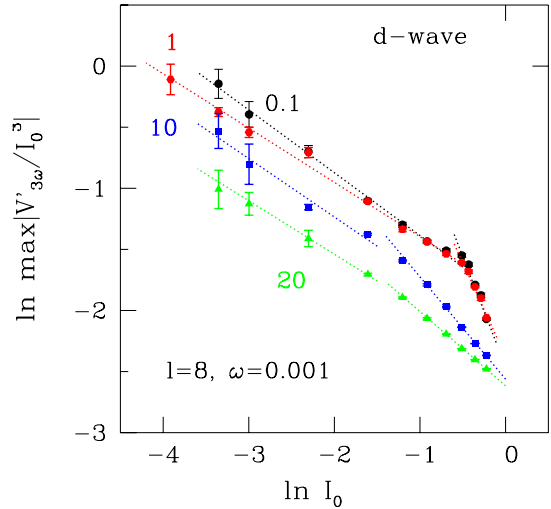}}
\caption{The current dependence of $max |V'_{3\omega}/I_0^3|$ for
$d$-wave system. We choose the system size $l=8$,
$\omega=0.001$ and $\tilde{L}$= 0.1, 1, 10 and 20 (its values are
shown next to the curves).
For each inductance
one has two distinct current regimes. 
The results are averaged over 10 - 800 samples.}
\end{figure}

In the experiment of Ref.\onlinecite{Matsuura}
the temperature $T_p$ is
merely an intergrain ordering transition temperature
above which the thermoremanent magnetization disappears.
In the previous simulations of \cite{LiDomin}
for the $d$-wave system, $T_p$  
is the temperature where there is an
onset of positive magnetization, i.e. the paramagnetic  Meissner effect
starts to be observed, but it does not seem to correspond to a
phase transition. The chiral glass phase transition temperature
$T_{cg}$ is found at
a lower temperature, $T_{cg}<T_p$ (for $L=1$, e.g., $T_{cg} \approx 0.29$
\cite{KawLi97}). 
Kawamura \cite{Kaw2000} 
has found that $z\approx 6 > 5$ for
the chiral glass transition, and thus no peak in $\rho_2(T)$ is expected
for this transition according to the scaling argument. 
Therefore, the peak measured by Yamao {\em et. al.}
may not correspond to the chiral glass transition, but to the 
crossover  we find at  $T_p$ for the $d$-wave case. 

In order to compare our results with experiments we first show that Yamao
{\em et. al.} \cite{Matsuura} performed measurements in the SCR.
Since real current is $I = \frac{2eJ}{\hbar}I_0$, $J \sim 10^2$ K and 
$I_0 \sim 10^{-1}$ we have $I \sim 10^{-2}$ mA. On the other hand,
the current used in experiments $I \sim 10$ mA  suggests that
the experiments were performed in the SCR.
As seen from Fig. 5, the value of $\alpha$ in the SCR for
$1 < \tilde{L} < 5$ coincides with the experimental value
\cite{Matsuura}. This interval of inductance is
realistic for ceramics \cite{Marcon} because typical
values of
$\tilde{L}$ are bigger than 3. An accurate comparison between theory and
experiments requires, however, the knowledge of 
$\tilde{L}$ which is not known for the compound of YBa$_2$Cu$_4$O$_8$ studied
in Ref. \onlinecite{Matsuura}. 

\begin{figure}
\epsfxsize=3.2in
\centerline{\epsffile{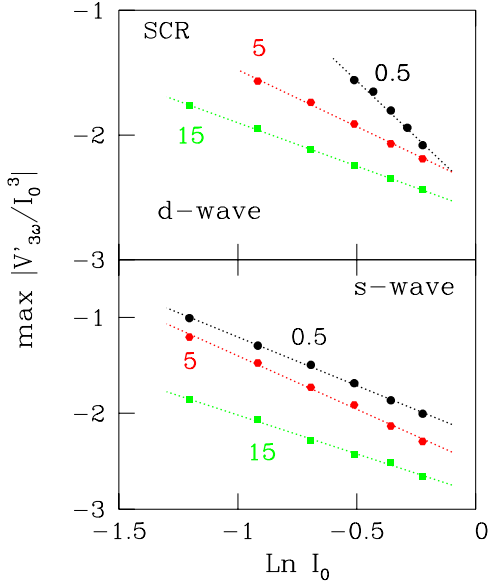}}
\vspace{0.5cm}
\caption{The current and self-inductance
 dependence of $max |V'_{3\omega}/I_0^3|$ for
$d$- and $s$-wave systems in the SCR for $\tilde{L}$=
0.5, 5 , and 15  
(they are shown next to the curves).
We choose the system size $l=8$ and
$\omega=0.001$.
The results are averaged over 5 - 10 samples.}
\end{figure}

\begin{figure}
\epsfxsize=3.2in
\centerline{\epsffile{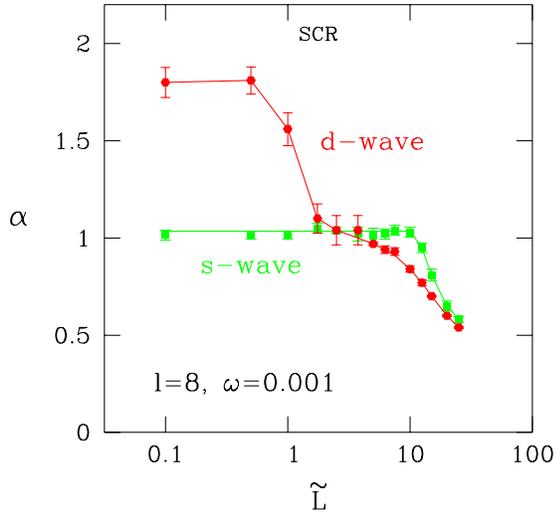}}
\caption{Dependence of $\alpha$ on $\tilde{L}$ in the
SCR for $s$- and $d$-wave systems.}
\end{figure}

In conclusion, we have calculated the non-linear ac resistivity 
exponent $\alpha$ for $s$ and $d$-wave
granular superconductors with high accuracy. Our results reveal two
distinct current regimes. In the WCR $\alpha$ is independent of
the screening strength and of types of pairing symmetry.
In the opposite case this exponent depends on $\tilde{L}$.
A difference between $s$- and $d$-wave symmetries in
the nonlinear resistivity can only be found in samples  with weak
screening.
The agreement between simulation and experimental results is possible
for some interval of $\tilde{L}$.

Financial support from the Polish agency KBN
(Grant No 2P03B-146-18), Conicet and ANPCYT
(Argentina)  and the Vietnam National Program on Basic Research
is acknowledged. We are very thankful to Dr. H. L. Minh and NCLAB
for prolonged use of their computational facility.
\par

\end{document}